\documentclass[eqsecnum,preprintnumbers,amsmath,amssymb]{revtex4}
\usepackage{epsfig,graphicx}% Include figure files
\usepackage{bm}% bold math
\begin{document}
\title{Mapping onto Electrodynamics of a\\
Two-dimensional 
Bose-Einstein Condensate}
\author{H. M. Cataldo}                     
\email{cataldo@df.uba.ar}
\homepage{http://www.df.uba.ar/users/cataldo}
\affiliation{IFIBA-CONICET\\
Departamento de F\'{\i}sica, Facultad de Ciencias Exactas y Naturales, \\
Universidad de Buenos Aires, RA-1428 Buenos Aires, Argentina}

\begin{abstract}
A thorough mapping between
the hydrodynamics of a two-dimensional Bose-Einstein condensate
and the nonrelativistic classical 
electrodynamics of a charged material medium 
is proposed. 
This is shown to provide a very useful frame 
to discuss several features of vortex dynamics. Two important
local conservation theorems of energy and angular momentum
are derived and further applications are summarized.
\end{abstract}
\maketitle
\section{Introduction}\label{s1}
Popov~\cite{popov} was the first to explain how the hydrodynamics 
of vortices in two-dimensional (2D) homogeneous superfluids
can be mapped onto (2+1)D relativistic electrodynamics, with vortices playing the role of
charged particles and phonons the role of photons.
The dynamics of superfluid films was later investigated by exploiting the analogy with
the dynamics of a 2D plasma~\cite{ambe}. 
The mapping onto (2+1)D electrodynamics was utilized in Ref.~\cite{arovas} to explore
the theory of dynamical vortices in superfluid films, deriving a frequency dependent
vortex mass. Such a mapping was also employed to study the phonon radiation by a vortex,
which exercises circular motion under the influence of an external force
 in a 2D homogeneous superfluid~\cite{lundh}. More recently, Fedichev~{\it et al.} \cite{fedi},
 using the analogy to electron-positron pair creation
in quantum electrodynamics,
investigated
the process of vortex-antivortex pair creation in a supersonically expanding and 
contracting quasi-2D Bose-Einstein condensate (BEC) at zero temperature.
However, in spite of the current interest and widespread 
importance of the phenomenon of Bose-Einstein
condensation of dilute gases in traps, the analogy of 2D BECs with
electrodynamics remains almost unexplored.

We propose in this paper a thorough mapping of the dynamics of a 2D BEC
onto the electrodynamics of a 2D material medium. 
After reviewing the main features of a 2D BEC in Sec.~\ref{sII}, we present such a mapping
in Sec.\ref{mapp}. Next, this is applied in Secs.~\ref{sIV} and \ref{sV} to deriving local conservation 
theorems of energy and angular momentum, respectively.  Finally, further 
applications are summarized in Sec.~\ref{conclusion}.

\section{Two-dimensional BEC}\label{sII}

Our starting point is the effective 2D Gross-Pitaevskii equation \cite{castin}
\begin{equation}
\label{gp}
i\hbar\frac{\partial}{\partial t}\Phi({\bf r},t)=\left(-\frac{\hbar^2\nabla^2}{2m}
+V_{\rm ext}({\bf r})+g|\Phi({\bf r},t)|^2\right)\Phi({\bf r},t),
\end{equation}
where $V_{\rm ext}({\bf r})$ denotes the trapping potential and $g$ corresponds to the effective
2D coupling constant between the atoms of mass $m$. The complex order parameter $\Phi$ may be
written in terms of a modulus and a phase, as follows:

\begin{equation}
\Phi({\bf r},t)=\sqrt{n({\bf r},t)}\exp[iS({\bf r},t)],
\end{equation}
where $n({\bf r},t)$ denotes the particle density and
the gradient of the phase $S$  yields the velocity field,

\begin{equation}
{\bf v}({\bf r},t)=\frac{\hbar}{m}{\bf \nabla}S({\bf r},t).
\end{equation}
Such a field is irrotational except at points ${\bf r}_j$, where the phase presents
a singularity corresponding to a quantized vortex, i.e.
\begin{equation}
\label{rot}
{\bf \nabla}\times{\bf v}=\sum_j \kappa_j\, \delta({\bf r}-{\bf r}_j)\,{\bf\hat{z}},
\end{equation}
$\kappa_j$ being the circulation of the velocity field along a  path encircling 
a single vortex located at ${\bf r}_j$.

The Gross-Pitaevskii equation (\ref{gp}) is equivalent to the following two 
coupled equations for the density and the velocity field \cite{dalf}:
\begin{equation}
\label{cont}
\frac{\partial n}{\partial t}=-{\bf \nabla}\cdot ({\bf v}n)
\end{equation}

\begin{equation}
\frac{\partial {\bf v}}{\partial t}=-{\bf \nabla}\left(\frac{V_{\rm ext}}{m}
+\frac{gn}{m}-\frac{\hbar^2}{2m^2\sqrt{n}}\nabla^2\sqrt{n}
+\frac{v^2}{2}\right).
\end{equation}
Assuming that the repulsive interaction among atoms is strong enough,
the density profiles are smooth and one can safely neglect the kinetic-pressure
term, proportional to $\hbar^2$, in the last equation, which then takes the form
\begin{equation}
\label{apo}
\frac{\partial {\bf v}}{\partial t}=-{\bf \nabla}\left(\frac{V_{\rm ext}}{m}
+\frac{gn}{m}+\frac{v^2}{2}\right).
\end{equation}
This result corresponds to the equation of potential flow for a fluid
whose pressure and density are related by the equation of state
\begin{equation}
\label{press}
p=g\,n^2/2.
\end{equation}
The approximation (\ref{apo}), however, overlooks the vortex dynamics represented by the time derivative of eq. 
(\ref{rot}), which may be explicitly taken into account by adding a Magnus force term \cite{sonin}:

\begin{equation}
\label{vel}
\frac{\partial {\bf v}}{\partial t}=-{\bf \nabla}\left(\frac{V_{\rm ext}}{m}
+\frac{gn}{m}+\frac{v^2}{2}\right)
+\sum_j \kappa_j\, \delta({\bf r}-{\bf r}_j)\,\dot{\bf  r}_j\times{\bf\hat{z}}.
\end{equation}
The equation of continuity (\ref{cont}) and the momentum equation
 (\ref{vel}) represent a system of hydrodynamic-type equations, which 
after linearization,
yields sound waves propagating at the local sound velocity
\begin{equation}
\label{cs}
c_s=\sqrt{\frac{1}{m}\frac{\partial p}{\partial n}}=\sqrt{\frac{gn}{m}}\,.
\end{equation}

Having explicitly introduced the vortex coordinates ${\bf  r}_j$ into the condensate equations of motion,
one must consider additional equations given by  vanishing Magnus forces acting upon each vortex:
\begin{equation}
\label{magnus}
n'({\bf  r}_j)\,m\,\kappa_j\,{\bf\hat{z}}\times[\dot{\bf  r}_j-{\bf  v}'({\bf  r}_j)]=0,
\end{equation}
where the prime on particle density and velocity field denotes that the
self-contribution of the corresponding vortex must be ignored.
Note that the above equation simply
states that the vortex core will move with the background superfluid velocity. However,
various effects causing possible departures from this behavior have been pointed out
\cite{arovas,nil,pra08,nova}.

\section{Mapping onto Electrodynamics\label{mapp}}

The above two-dimensional dynamics can be mapped onto the electrodynamics of a
 2D material medium, with electromagnetic waves representing sound waves and
macroscopic (free) point charges representing vortices.
We shall utilize in the following the Heaviside-Lorentz system of units \cite{jackson}. 
To begin with, we assume a transverse magnetic induction proportional to the condensate
density
\begin{equation}
\label{b}
{\bf B}=\sqrt{g}\,n\,{\bf\hat{z}}
\end{equation}
and an electric field, whose expression arises from representing vortices as point charges
of negligible mass, leading to a vanishing Lorentz force
\begin{equation}
\label{lor}
{\bf E}+\frac{{\bf v}}{c}\times{\bf B}=0,
\end{equation}
where $c=\sqrt{g\,n_{\rm max}/m}$ denotes the maximum speed of sound in the condensate.
The above expression, rewritten as $E/B=v/c$, embodies an important analogy to quantum electrodynamics.
In fact, the Schwinger vacuum breakdown \cite{schwinger} is a phenomenon occuring whenever the 
electric field exceeds the magnetic field, which gives rise to electron-positron pair creation.
On the other hand, it is well known that a supersonic superflow becomes unstable with
respect to the spontaneous creation of vortex-antivortex pairs \cite{fedi}.

We shall restrict ourselves to slow motion superflows $v/c\ll 1$,
which ensures a mapping onto nonrelativistic classical electrodynamics.
Then, from eqs. (\ref{b}) and (\ref{lor}) we have
\begin{equation}
\label{e}
{\bf E}=\frac{\sqrt{g}}{c}\,n\,{\bf\hat{z}}\times{\bf v}.
\end{equation}
Equation (\ref{b}) implies the absence of free magnetic poles, 
${\bf \nabla}\cdot{\bf B}=0$, while the continuity equation 
(\ref{cont}) becomes equivalent to Faraday's law:
\begin{equation}
\label{far}
{\bf \nabla}\times{\bf E}=-\frac{1}{c}\frac{\partial{\bf B}}{\partial t}.
\end{equation}
The electric displacement field may be obtained by assuming a permittivity given by,
\begin{equation}
\label{eps}
\epsilon=\frac{c^2}{c_s^2}=\frac{n_{\rm max}}{n}
\end{equation}
and therefore
\begin{equation}
\label{d}
{\bf D}=\epsilon\,{\bf E}=\sqrt{m\,n_{\rm max}}\,\,\,{\bf\hat{z}}\times{\bf v}.
\end{equation}
Thus, the curl equation (\ref{rot}) yields Coulomb's law:
\begin{equation}
\label{coulomb}
{\bf \nabla}\cdot{\bf D}=\rho=\sum_j q_j\, \delta({\bf r}-{\bf r}_j),
\end{equation}
where a negative charge corresponds to a positive vorticity and viceversa,
\begin{equation}
\label{charge}
q_j=-\,\sqrt{m\,n_{\rm max}}\,\,\kappa_j.
\end{equation}
Particularly, the velocity field of the vortex located at ${\bf r}_j$ reads
\begin{equation}
\label{campo}
{\bf v}_j=\frac{\kappa_j}{2\pi}\,\,{\bf\hat{z}}\times\frac{({\bf r}-{\bf r}_j)}
{|{\bf r}-{\bf r}_j|^2},
\end{equation}
which, according to (\ref{d}) and  (\ref{charge}), yields
\begin{equation}
{\bf D}_j=\frac{q_j}{2\pi}\,\,\frac{({\bf r}-{\bf r}_j)}
{|{\bf r}-{\bf r}_j|^2},
\end{equation}
that is, the correct expression for the field of a point charge $q_j$ in two dimensions.

Finally, the momentum equation (\ref{vel}) yields Amp\`ere-Maxwell law,
\begin{equation}
\label{amp}
{\bf \nabla}\times{\bf H}-{\bf J}/c = 
\frac{1}{c}\frac{\partial {\bf D}}{\partial t},
\end{equation}
where the current density is given by
\begin{equation}
\label{current}
{\bf J}=\sum_j q_j\, \delta({\bf r}-{\bf r}_j)\,\dot{{\bf r}}_j,
\end{equation}
which is in agreement with the charge density of the right-hand side of eq.~(\ref{coulomb}).
The magnetic field is ${\bf H}={\bf B}-{\bf M}$ with a magnetization given by
\begin{equation}
\label{mag}
{\bf M}=-\left(\frac{V_{\rm ext}}{\sqrt{g}}+\frac{\sqrt{g}\,D^2}{2\,m\,c^2}\right)
\,{\bf\hat{z}},
\end{equation}
that is, a permanent magnet perturbed by a constitutive second-order correction
in the displacement field. 

The above Maxwell equations
must be suplemented with vanishing Lorentz forces acting upon each point charge
(cf. eqs. (\ref{magnus}) and (\ref{lor})),
\begin{equation}
\label{lorentz}
q_j\left[{\bf E}'({\bf r}_j)+\frac{\dot{{\bf r}}_j}{c}\times{\bf B}'({\bf r}_j)\right]=0,
\end{equation}
where the primed fields denote that they are excluding the self-contribution of the 
charge $q_j$.

\section{Conservation of Energy}\label{sIV}

Now we shall analyze Poynting's
theorem \cite{jackson}. The total rate of doing work by the fields is ${\bf J}\cdot{\bf E}$
per unit area. This power must be balanced by a corresponding rate of decrease of energy in
the electromagnetic field. In order to exhibit this conservation law explicitly, we first use
the Amp\`ere-Maxwell law (\ref{amp}) to eliminate ${\bf J}$,
\begin{equation}
\label{JE}
{\bf J}\cdot{\bf E}=c\,{\bf E}\cdot{\bf \nabla}\times{\bf H}-{\bf E}\cdot\frac{\partial{\bf D}}
{\partial t}.
\end{equation}
Now, employing the vector identity ${\bf E}\cdot{\bf \nabla}\times{\bf H}={\bf H}\cdot{\bf \nabla}\times{\bf E}
-{\bf \nabla}\cdot({\bf E}\times{\bf H})$ and Faraday's law (\ref{far}), the right side of (\ref{JE}) becomes
\begin{eqnarray}
\label{JE1}
{\bf J}\cdot{\bf E}&=&-{\bf H}\cdot\frac{\partial{\bf B}}{\partial t}
-{\bf E}\cdot\frac{\partial{\bf D}}{\partial t}-c\,
{\bf \nabla}\cdot({\bf E}\times{\bf H})\nonumber\\
&=&-\frac{\partial}{\partial t}({\bf B}\cdot{\bf H}-B^2/2)
-c\,{\bf \nabla}\cdot({\bf E}\times{\bf H}),
\end{eqnarray}
where the last line stems from (\ref{mag}) and (\ref{d}).
Here we may recognize the differential form of Poynting's theorem
\begin{equation}
\label{po}
\frac{\partial u}
{\partial t}+
{\bf \nabla}\cdot{\bf S}=-{\bf J}\cdot{\bf E}
\end{equation}
with
\begin{equation}
\label{u}
u={\bf B}\cdot{\bf H}-B^2/2=n\left(\frac{mv^2}{2}+V_{\rm ext}+\frac{gn}{2}\right)
\end{equation}
the condensate energy density, and the vector
\begin{equation}
\label{S}
{\bf S}=c\,\,{\bf E}\times{\bf H}=\frac{\partial u}{\partial n}\,n\,{\bf v},
\end{equation}
representing energy flow, which may be called the BEC Poynting's vector.

Taking into account (\ref{current}), the right side of eq.~(\ref{po}) reads
\begin{equation}
\label{work}
-{\bf J}\cdot{\bf E}=-\sum_j q_j\, \delta({\bf r}-{\bf r}_j)\,{\bf E}'({\bf r}_j)\cdot\dot{{\bf r}}_j,
\end{equation}
where we have disregarded the self-field contribution to the work. Two important conclusions may be 
drawn from the above expression. First, by comparing eqs. (\ref{lorentz}) and (\ref{work}), we may see
that Poynting's  theorem for the condensate simply reads
\begin{equation}
\label{poyn}
\frac{\partial u}
{\partial t}+
{\bf \nabla}\cdot{\bf S}=0,
\end{equation}
which means that energy can only change through the flow of Poynting's vector. The second important expression
arises as follows. From (\ref{po}) and (\ref{work}) we have
\begin{equation}
\label{gradu}
\frac{\partial u}{\partial {\bf r}_j}=- q_j\, \delta({\bf r}-{\bf r}_j)\,{\bf E}'({\bf r}_j),
\end{equation}
which integrated over the condensate yields,
\begin{equation}
\label{e'}
{\bf E}'({\bf r}_j)=-\frac{1}{q_j}\frac{\partial U}{\partial {\bf r}_j},
\end{equation}
being $U$ the condensate energy. Then, inserting the above expression into eq. (\ref{lorentz})
we obtain
\begin{equation}
\label{velvor}
\dot{{\bf r}}_j=\frac{1}{m\,\kappa_j\, n_0({\bf r}_j)}\,\,\frac{\partial U}{\partial {\bf r}_j}\times
{\bf\hat{z}},
\end{equation}
where the condensate density in absence of the vortex, $n'({\bf r}_j)$, was approximated by the 
density of the ground state $n_0({\bf r}_j)$. Expression (\ref{velvor}) has proven very useful for
numerically evaluating the vortex velocity in terms of the condensate energy gradient, yielding
accurate results even in cases where the vortex velocity differs appreciably from the background
superfluid velocity \cite{laser,epjd}.

\section{Conservation of Angular Momentum}\label{sV}
In this section we shall utilize the electromagnetic analogy to
derive a local conservation law of angular momentum for a BEC in an axisymmetric
trap. 
First, using the Maxwell equations, we may write the electromagnetic force per unit area
as \cite{jackson},
\begin{equation}
\label{eforce}
\rho\,{\bf E}+\frac{1}{c}\,{\bf J}\times{\bf B}={\bf E}\cdot{\bf \nabla}\,{\bf D}-{\bf D}\times
({\bf \nabla}\times{\bf E})-{\bf B}\times({\bf \nabla}\times{\bf H})
-\frac{1}{c}\,\frac{\partial}{\partial t}\,({\bf D}\times{\bf B}).
\end{equation}
Then, taking into account the axial symmetry of the trapping potential, the torque per unit 
area may be written
\begin{eqnarray}
\label{etorque}
{\bf r}\times\left(\rho\,{\bf E}+\frac{1}{c}\,{\bf J}\times{\bf B}\right)&
= & {\bf\hat{z}}\left\{x\left[\frac{\partial}{\partial y}(D_y^2/\epsilon)+
\frac{\partial}{\partial x}(D_xD_y/\epsilon)-
\frac{\partial}{\partial y}(D^2/\epsilon)-
\frac{\partial}{\partial y}(B^2/2)\right]\right.\nonumber\\
&&-
\left.y\left[\frac{\partial}{\partial x}(D_x^2/\epsilon)+
\frac{\partial}{\partial y}(D_xD_y/\epsilon)-
\frac{\partial}{\partial x}(D^2/\epsilon)-
\frac{\partial}{\partial x}(B^2/2)\right]\right\}\nonumber\\
&&-\frac{1}{c}\,\frac{\partial}{\partial t}\,[{\bf r}\times({\bf D}\times{\bf B})].
\end{eqnarray}
On the other hand, from Newton's second law, such a torque corresponds to 
the rate of change of the density of mechanical angular momentum ${\bf \cal L}_{\rm mech}$.
Therefore, eq. (\ref{etorque}) may be rewritten as the differential form of the conservation
law of angular momentum:
\begin{equation}
\label{consl}
\frac{\partial}{\partial t}({\bf \cal L}_{\rm mech}+{\bf \cal L}_{\rm field})
+{\bf \nabla}\cdot{\bf \cal M}=0,
\end{equation}
where the field angular-momentum density is
\begin{equation}
\label{lfield}
{\bf \cal L}_{\rm field}=
{\bf r}\times({\bf D}\times{\bf B})/c=n\,{\bf r}\times(m{\bf v}),
\end{equation}
i.e., the BEC angular-momentum density.
The flux of angular momentum in (\ref{consl}) is described by the pseudo tensor
\cite{jackson}
\begin{equation}
\label{pseu}
{\bf \cal M}=
{\bf \cal T}\times\,{\bf r},
\end{equation}
where the tensor ${\bf \cal T}$ has nonvanishing components only in the $x$-$y$ subspace
\begin{eqnarray}
\label{compt}
{\cal T}_{ij} &=& D_iD_j/\epsilon-\delta_{ij}(D^2/\epsilon+B^2/2)\nonumber\\
&=& -m\,n\, v_i\, v_j - \delta_{ij}\,p,
\end{eqnarray}
with the pressure $p$ given by eq.~(\ref{press}).
Here we may recognize $-{\bf \cal T}$ as the BEC momentum flux density tensor.

Next we show that the torque (\ref{etorque}) vanishes. To this aim, we may rewrite the
force given by the left side of (\ref{eforce}) using eqs.~(\ref{coulomb}) and (\ref{current}),
which yields
\begin{equation}
\label{force0}
\sum_j \delta({\bf r}-{\bf r}_j)\,\,q_j
\left[{\bf E}'({\bf r}_j)+\frac{\dot{{\bf r}}_j}{c}\times{\bf B}'({\bf r}_j)\right],
\end{equation}
where we have excluded self-force contributions. But, according to (\ref{lorentz}),
 the above expression between square brackets
must vanish, which yields a vanishing torque.
Hence, the BEC conservation law of angular momentum finally reads
\begin{equation}
\label{consll}
\frac{\partial}{\partial t}(n\,m\,{\bf r}\times{\bf v})
+{\bf \nabla}\cdot{\bf \cal M}=0,
\end{equation}
which means that angular momentum can only change through the
flow of the tensor ${\bf \cal M}$.

\section{Further applications\label{conclusion}}

The following applications of the present formalism are treated in Ref.~\cite{nova}:
\begin{itemize}
\item The velocity field induced by an off-centered vortex, which is analyzed
 from the viewpoint of the equivalent
electromagnetic problem. 
\item Two intriguing aspects of the vortex dynamics:
\begin{itemize}
\item
The recently found 
`core effect' on the vortex velocity \cite{nil,pra08}, which is
 interpreted within the electromagnetic picture.
\item
The highly controversial question of the vortex inertia, which is discussed from
the viewpoint of the electrodynamic analogy.
\end{itemize}
\end{itemize}

\end{document}